
%
%
%
%
%
%
%
%
%

\input harvmac

\overfullrule=0mm

\def\tab{\Theta_{a,b}}
\def\fab{f_{a,b}}
\def\zt{{z|\tau}}
\def\hlf{{1\over 2}}
\def\ip{{i\pi }}
\def\pab{\Phi_{a,b}}
\def\chat{{\hat c}}

\def\encadre#1{\vbox{\hrule\hbox{\vrule\kern8pt\vbox{\kern8pt#1\kern8pt}
\kern8pt\vrule}\hrule}}
\def\encadremath#1{\vbox{\hrule\hbox{\vrule\kern8pt\vbox{\kern8pt
\hbox{$\displaystyle #1$}\kern8pt}
\kern8pt\vrule}\hrule}}

%
\def\frac#1#2{\scriptstyle{#1 \over #2}}

%
%

%
\def\({ \left( }
\def\){ \right) }
%


\def\IR{\relax{\rm I\kern-.18em R}}
\font\cmss=cmss10 \font\cmsss=cmss10 at 7pt
\def\IZ{\relax\ifmmode\mathchoice
{\hbox{\cmss Z\kern-.4em Z}}{\hbox{\cmss Z\kern-.4em Z}}
{\lower.9pt\hbox{\cmsss Z\kern-.4em Z}}
{\lower1.2pt\hbox{\cmsss Z\kern-.4em Z}}\else{\cmss Z\kern-.4em Z}\fi}
\def\inbar{\,\vrule height1.5ex width.4pt depth0pt}
\def\IB{\relax{\rm I\kern-.18em B}}
\def\IC{\relax\hbox{$\inbar\kern-.3em{\rm C}$}}
\def\ID{\relax{\rm I\kern-.18em D}}
\def\IE{\relax{\rm I\kern-.18em E}}
\def\IF{\relax{\rm I\kern-.18em F}}
\def\IG{\relax\hbox{$\inbar\kern-.3em{\rm G}$}}
\def\IH{\relax{\rm I\kern-.18em H}}
\def\II{\relax{\rm I\kern-.18em I}}
\def\IK{\relax{\rm I\kern-.18em K}}
\def\IL{\relax{\rm I\kern-.18em L}}
\def\IM{\relax{\rm I\kern-.18em M}}
\def\IN{\relax{\rm I\kern-.18em N}}
\def\IO{\relax\hbox{$\inbar\kern-.3em{\rm O}$}}
\def\IP{\relax{\rm I\kern-.18em P}}
\def\IQ{\relax\hbox{$\inbar\kern-.3em{\rm Q}$}}
\def\IGa{\relax\hbox{${\rm I}\kern-.18em\Gamma$}}
\def\IPi{\relax\hbox{${\rm I}\kern-.18em\Pi$}}
\def\ITh{\relax\hbox{$\inbar\kern-.3em\Theta$}}
\def\IOm{\relax\hbox{$\inbar\kern-3.00pt\Omega$}}





\def\Gl{\lambda}


%
%
%
%
%
\Title{SPhT 93/068--TAUP 2069-93 hep-th/9306157}
{{\vbox {\vskip-.5in
\centerline{Elliptic Genera and the Landau--Ginzburg}
\bigskip
\centerline{Approach to  N=2  Orbifolds}
}}}
\centerline{P. Di Francesco,}
\medskip\centerline
{\it Service de Physique Th\'eorique de Saclay
\footnote{$^{\#}$}{
Laboratoire de la Direction des Sciences et de la Mati\`ere
du Commissariat \`a l'Energie Atomique},}
\centerline{\it 91191 Gif sur Yvette Cedex, France,}
\medskip
\centerline{}
\medskip
\centerline{O. Aharony and S. Yankielowicz \footnote{$^{\$}$}{Work supported
in part by the US--Israel Binational Science Foundation
and the Israel Academy of Sciences.},}
\medskip\centerline{\it School of Physics and Astronomy,}
\centerline{\it Beverly and Raymond Sackler Faculty of Exact
Sciences,}
\centerline{\it Tel--Aviv University,
Ramat--Aviv, Tel--Aviv 69978, Israel.}
\vskip .2in
We compute the elliptic genera of orbifolds associated with $N=2$
super--conformal theories which admit a Landau-Ginzburg description. The
identification
of the elliptic genera of the macroscopic Landau-Ginzburg orbifolds with those
of the corresponding microscopic $N=2$ orbifolds further supports the
conjectured
identification of these theories. For $SU(N)$ Kazama-Suzuki models the
orbifolds
are associated with certain $\IZ_p$ subgroups of the various coset
factors. Based
on our approach we also conjecture the existence of "$E$-type"
variants of these
theories, their elliptic genera and the corresponding Landau-Ginzburg
potentials.

\Date{06/93}
%
\newsec{Introduction}

The proposal of an effective field theoretical description
of $N=2$ superconformal theories ``\`a la Landau--Ginzburg" (LG)
has recently received much further support.
On the one hand,
one is dealing with the now familiar $N=2$ rational superconformal
field theories, mainly obtained by a coset construction  using
WZW factors. In the following we will mainly concentrate on
the Kazama--Suzuki (KS) coset models of the form
$SU(N)_k\times SO(2(N-1))_1/SU(N-1)_{k+1} \times U(1)$, which will be
referred to as ``$SU(N)$ KS models".
A tentative classification of such theories is made possible
by the knowledge of modular invariants for the various WZW factors.
Unfortunately only the $SU(2)$
\ref\SUSU{A. Cappelli, C. Itzykson and J.-B. Zuber, Nucl. Phys.
{\bf B280 [FS18]} (1987) 445; Comm. Math. Phys. {\bf 113} (1987) 1.}
and $SU(3)$
\ref\SU{P. Christe and F. Ravanini,
Int. J. Mod. Phys. {\bf A4} (1989) 897;D. Altsch\"uler,
J. Lacki and P. Zaugg, Phys. Lett. {\bf B205} (1988) 281.;
        P. Ruelle, E Thiran and J. Weyers,
{\it Implications of an arithmetical
symmetry of the commutant for modular invariants}, DIAS preprint
STP--92--26 (1992); T. Gannon,
{\it The Classification of SU(3) modular
invariant partition functions}, Carleton preprint
92-061 (1992);
T. Gannon and Q. Ho-Kim, {\it The low level modular invariant
partition functions of rank two algebras}, Carleton
preprint 93-0364 (1993).}\
cases are completely
classified so far, and it remains a formidable task to go beyond them
(see however \ref\VAF{S. Cecotti and C. Vafa,
{\it On Classification of N=2 Supersymmetric Theories},
Harvard and Trieste preprint HUTP-92/A064, SISSA-203/92/EP (1992).}\
for some alternative route).
However, the orbifold procedure enables us to manufacture non--trivial
modular invariant theories, by moding out some symmetry of the
initial WZW theory. The simplest of these is the
$\IZ_p$ orbifold of $SU(N)$ WZW theories, for $p$ a divisor of $N$.

On the other hand, we have an effective description of such theories
using some $N=2$ superfields $\Phi_i$ governed by the
Landau-Ginzburg action
$$ S= \int d^2x d^4 \theta \ \Phi_i {\bar \Phi_i}
+\{ \int d^2x d^2 \theta \ {W(\Phi_i)} + {\rm c.c.}\},$$
where $W$ is some quasi--homogeneous polynomial potential of the
graded fields $\Phi_i$.
In addition to the identification of central charges of
both theories
\ref\ACT0{C. Vafa and N. Warner, Phys. Lett. {\bf B218} (1989) 51.}
(linked to the degrees of $W$ and of the fields $\Phi_i$),
the ``chiral rings" of the
superconformal theories
(the ring structure formed by the chiral primary fields under OPE)
were also identified
with those of the corresponding LG descriptions
\ref\EMLG{D. Gepner, Comm. Math. Phys. {\bf 141} (1991) 381 ;
          W. Lerche, C. Vafa and N. Warner, Nucl. Phys.
{\bf B324} (1989) 427;
S. Cecotti, L. Girardello and A. Pasquinucci, Nucl. Phys. {\bf B238}
(1989) 701 and Int. J. Mod. Phys. {\bf A6} (1991) 2427;
S. Cecotti, Int. J. Mod. Phys. {\bf A6} (1991) 1749. }
(the ground state ring
$\IC[x_1,x_2,...] / \nabla W$). Recently E. Witten
\ref\WIT{E. Witten,
{\it On the Landau-Ginzburg description of
N=2 minimal models,} preprint IASSNS--HEP--93/10 (1993).}\
was able to compute
yet another quantity in the LG framework for the $SU(2)$ KS models of
diagonal $A$-type modular invariant,
and to compare it with the corresponding superconformal field theory
results.
This quantity is the elliptic genus, defined as a certain twisted
boundary condition toroidal partition function
\eqn\ellipgen{ Z_2(u|\tau)={\sum_{l}}'
{\rm Tr}_{{\cal R}_{l}} (-1)^{F} q^{H_L}
e^{i \gamma_L J_{0,L}}, }
where $F$, $H_L$, $J_{0,L}$ and $\gamma_L$ denote respectively
the total fermion number, the hamiltonian
$H_L=L_0-{c \over 24}$, the zero mode of
the $U(1)$ symmetry generator
and the associated charge of the left--moving Ramond states.
The sum extends over the states with
vanishing right--moving hamiltonian         and $U(1)$ charge,
$H_R=\gamma_R=0$, which are
Ramond sector representations of the $N=2$ superalgebra containing
a ground state of $H_L=0$.
In the following, we denote $u={\gamma_L \over 2 \pi (k+N)}$
and $q=e^{2i \pi \tau}$ (for a general LG theory the degree of $W$ will
appear in the expression for $u$ instead of $k+N$).
The direct LG calculation of this quantity turns out to be
a simple free field calculation : due to the topological
invariance of \ellipgen \WIT\ , one can set the potential piece
of the action to zero, after integration over the bosonic
upper components of the superfields,
without altering the result. The actual calculation only involves
sorting out the left/right moving bosonic and fermionic mode
contributions, and finally gives rise to a product
formula for the LG elliptic genus.

In the $N=2$ superconformal framework, the elliptic genus is expressed as
a sum over (twisted) Ramond characters. In
\ref\DY{P. Di Francesco and S. Yankielowicz,
Saclay and Tel--Aviv University preprint SPhT 93/049 TAUP 2047-93.},
two of us established a general scheme for proving the identity between
the elliptic genera in the two approaches, and were able to extend
this to the $D$ and $E$--type modular invariant $SU(2)$ KS theories, as
well as to the $A$--type invariants of the $SU(N)$ KS models.
The proof is based on an elementary lemma on elliptic modular functions.
The idea is to show  that both elliptic genera behave similarly under
the elliptic $z \to z+1$, $z \to z+\tau$ and modular
$\tau \to \tau +1$ and $(z,\tau) \to ({z \over \tau},-{1 \over \tau})$
transformations, and share the same $q \to 0$ (i.e. $\tau \to i \infty$)
limit. Consequently, the lemma states that their ratio is an elliptic modular
form of weight zero, with limit $1$ when $q \to 0$.
Therefore, it is equal to $1$ identically.
Usually the first two transformations are easy to find, using the definition
of the Ramond characters. The third one is obvious, due to the
identity for the character with highest representation of weight $h$
$$ \chi_h(z|\tau+1)=e^{2i \pi (h-{c \over 24})} \chi_h(z|\tau),$$
and since we have $h={c \over 24}$ for the Ramond states contributing to
the elliptic genus. The only tricky point is the last transformation.
We give now a general proof of modular
covariance of the $N=2$ superconformal elliptic genus.
First, the elliptic genus \ellipgen\ for a general $N=2$ superconformal
theory can be expressed as a particular limit of the modular covariant
(twisted) Ramond sector partition function
\eqn\partfr{\eqalign{{\cal Z}(z|\tau)&=\sum_{h,{\bar h}} N_{h,{\bar h}}
\chi_h(z|\tau) \chi_{\bar h}^*(z |\tau) \cr
&\equiv \phi(z,\tau,{\bar z}, {\bar \tau}),\cr}}
$N_{h,{\bar h}} \in \IN$, considered as a function of $z$ and $\bar z$
as independent variables, and with
$$\chi_h(z|\tau)={\rm tr}_{R_h}(q^{L_{0,L}-{c \over 24}}
e^{2i \pi \alpha J_{0,L} z} (-1)^F_L)$$
where $\alpha$ is a fixed factor which makes all $U(1)$ charges (eigenvalues
of $J_0$) integer: for the $SU(N)$ KS models, we have $\alpha=N(k+N)$,
and we have to substitute $h \to {\bar h}$, $L \to R$ for right moving
represenrtations.
Taking ${\bar z}=0$ in \partfr, we end up with the elliptic genus
\eqn\sumk{\phi(z,\tau,{\bar z}=0,{\bar \tau})=
K(z|\tau)={\sum_h}' \chi_h(z|\tau),}
where the sum extends over the same states as in eqn.\ellipgen
\foot{The notation $(h,{\bar h})$ for the left--right representations
involved is abusive, as $h={\bar h}={c \over 24}$ for all the remaining
states,
after one takes ${\bar z} \to 0$. What distinguishes them is actually
their U(1) charge, and the sum in \sumk\ is over the ``chiral" Ramond states,
obtained from the spinless ($h={\bar h}$) chiral states of the Neveu--Schwarz
sector by spectral flow. For instance in the SU(2) A, D, E cases, these
states are indexed by the Coxeter exponents (shifted by $-1$) of the
corresponding Lie algebra.}.
The modular covariance of the full partition function
\eqn\transmod{
{\cal Z}({z \over \tau}|-{1 \over \tau})=
e^{i \pi \beta ({z^2 \over \tau}-{{\bar z}^2 \over {\bar \tau}})}
{\cal Z}(z|\tau)}
follows from the simple coset form of the modular transformations of
the characters, explicitly factorizing a phase
$\exp(i \pi \beta ({z^2 \over \tau}-{{\bar z}^2 \over {\bar \tau}}))$,
with $\beta=N^2 (N-1)k(k+N)$ for the $SU(N)$ KS case.
Taking ${\bar z}=0$ in \transmod, we find that the elliptic genus of the
$N=2$ superconformal theory is a modular form of weight $0$, namely
\eqn\transell{
K({z \over \tau}|-{1 \over \tau})=e^{i \pi \beta  {z^2 \over \tau}}
K(z|\tau).}
Thus, whenever we will be able to compute a LG elliptic genus, our
first task will be to compute its elliptic and modular properties
and to compare its $q \to 0$ limit to that of the corresponding
$N=2$ superconformal one.

In this paper,  we address orbifolds within the LG approach
\ref\VAFORB{C. Vafa, Mod. Phys. Lett. {\bf A4} (1989), 1169.},
in particular those associated with the $SU(N)$ KS theories.
In the first section we compute the LG elliptic genus for
the $D$--type $\IZ_2$ orbifolds of the $SU(2)$ KS models,
giving rise to yet another identity between Jacobi theta functions.
In the second section, we address the $SU(3)$ case in detail:
in addition to the expected $\IZ_3$ orbifold, we can also carry out
a $\IZ_2$ orbifold corresponding to the $SU(2)$ factor of the
KS coset model. Modular invariance tells us also that some $E_{6,7,8}$
theories of the $SU(2)$ factor should also exist, and we give a natural
conjecture for their elliptic genera and LG potentials.
Next we turn to the general $\IZ_p$ orbifold case of $SU(N)$ KS models.
Although no direct LG description exists for the orbifolds in general,
the elliptic genera can be computed by twisted field techniques, starting
from the LG theory for $SU(N)$ KS models.
In the last section we analyze the elliptic genus of orbifolds of general
LG theories. We end with conclusions, remarks and open questions.

\newsec{The $SU(2)$ case: $D$ series as $\IZ_2$ orbifolds.}

The $N=2$ LG potentials for the $A_{k+1}$--type $SU(2)$ KS models
(i.e. the cosets ${SU(2)_k \over U(1)}\times U(1)$) read
\eqn\initpot{W_{k+2}^{(2,A)}(\Phi) = {\Phi^{k+2} \over k+2}.}
Whenever $k$ is even, say $k=2n$, the action has a $\Phi \to -\Phi$
symmetry. Moding out by this symmetry amounts to performing the
change of variable $\Phi_1=\Phi^2$ in the path integral\foot{
See appendix A of
ref.\ref\DLZ{P. Di Francesco, F. Lesage and J.-B. Zuber, {\it Graph Rings
and Integrable Perturbations of N=2 Superconformal Theories}, Saclay
preprint SPhT 93/057 (1993).}\
for a complete argument.}.
The resulting
Jacobian, $\Phi_1^{-1/2}$, can be put back into the action by introducing
a second field $\Phi_2$, finally leading to the effective potential
\ref\ACT{C. Vafa and N. Warner, Phys. Lett. {\bf B218} (1989) 51;
E. Martinec, Phys. Lett. {\bf B217} (1989) 431;
{\it Criticality, Catastrophes
and Compactifications}, Physics and Mathematics of Strings,
ed. L. Brink, D. Friedan and A. M. Polyakov (World Scientific, 1990).}
\eqn\secondd{
W_{2n+2}^{(2,D)}(\Phi_1,\Phi_2)={\Phi_1^{n+1} \over 2(n+1)}
+{1 \over 2} \Phi_1 \Phi_2^2.}
In \DY, we performed the direct free field calculation of the elliptic genus
starting from the latter LG potential.
The main point here is that the calculation can be done directly using
the initial potential \initpot. We proceed as for the calculation of the
one loop orbifold partition functions of free fields.
The 'R'--symmetry of the fields which preserves the action after
integration over the upper component $F$ of the superfield
$\Phi=\phi+\theta_+ \psi_+ + \theta_- \psi_-+\theta_+\theta_- F$
is
$$\eqalign{ \phi &\to e^{2i \pi u} \phi \cr
\psi_+ &\to e^{2i \pi u} \psi_+ \cr
\psi_- &\to e^{-2i \pi (k+1) u} \psi_- .\cr}$$
The additional $\IZ_2$ symmetry allowed by the potential \initpot\
for $k=2n$ translates into the possibility of an extra
{\it simultaneous} change
$$\phi \to - \phi \qquad \psi_{\pm} \to - \psi_{\pm}.$$
Including this possibility in the free field toroidal calculation of \WIT\
gives rise to four possible sectors (PP), (AP), (PA),
(AA), according to the sign chosen
in the $1$ or $\tau$ directions of the torus (e.g. the (AP) sector
corresponds to $\Phi(z+1) \to -\Phi(z)$ and
$\Phi(z+\tau) \to \Phi(z)$). The (PP) contribution is just that of the
non--twisted superfield, and coincides with the $A$--type theory result
\WIT\ \DY\
$$\eqalign{
Z_{PP}(u|\tau)&= {\Theta_1((k+1)u|\tau) \over \Theta_1(u|\tau)}\cr
&=e^{-i\pi k u}{1-e^{2i \pi (k+1)u}\over 1-e^{2i \pi u}}
\prod_{n \geq 1} {(1-q^ne^{2i \pi (k+1)u})(1-q^ne^{-2i \pi (k+1)u})
\over (1-q^ne^{2i \pi u})(1-q^ne^{-2i \pi u})},}$$
where we identify the contributions of left moving fermion
and boson modes of the free superfield.
Here and in the following, $\Theta_j(u|\tau)$, $j=1,2,3,4$, denote
the Jacobi theta functions.
The effect of the twist is clear: it amounts to
replacing $n$ by $n-{1 \over 2}$ (the modes of the
components of the superfield $\phi_n$ get twisted, i.e. changed into
$\phi_{n \pm {1 \over 2}}$ for antiperiodicity in the $1$ direction)
or $u \to u-{1 \over 2}$ (twist in the $\tau$ direction).
This gives
$$\eqalign{Z_{AP}(u|\tau)&=
{\Theta_4((k+1)u|\tau) \over \Theta_4(u|\tau)}\cr
Z_{PA}(u|\tau)&=
{\Theta_2((k+1)u|\tau) \over \Theta_2(u|\tau)}\cr
Z_{AA}(u|\tau)&=
{\Theta_3((k+1)u|\tau) \over \Theta_3(u|\tau)}.\cr}$$
    We get the final expression for the orbifold elliptic genus
\eqn\dseries{\eqalign{
Z_{2p+2}^{(D)}(u|\tau)&={1 \over 2}(Z_{PP}+Z_{AP}+Z_{PA}+Z_{AA})\cr
&=\hlf \sum_{j=1}^4 {\Theta_j((k+1)u|\tau) \over \Theta_j(u|\tau)}.\cr}}
A few remarks are in order. The above expression is a modular form
of weight zero, with simple $u \to u+2$ and $u\to u+2\tau$
transformations
$$\eqalign{
Z_{2p+2}^{(D)}(u+2|\tau)&= Z_{2p+2}^{(D)}(u|\tau)\cr
Z_{2p+2}^{(D)}(u+2\tau|\tau)&= e^{-4i \pi k(k+2)(u+\tau)}
Z_{2p+2}^{(D)}(u|\tau)\cr
Z_{2p+2}^{(D)}(u|\tau+1)&= Z_{2p+2}^{(D)}(u|\tau) \cr
Z_{2p+2}^{(D)}({u \over \tau}|-{1 \over \tau})
&= e^{i \pi k(k+2) {u^2 \over \tau}} Z_{2p+2}^{(D)}(u|\tau).\cr}$$
Note that this is true only for even $k$'s, otherwise some phases
would appear.
Moreover, its $q \to 0$ ($\tau\to i \infty$)
limit is easily derived, using the limits
$$\eqalign{
\Theta_1(u|\tau) \to 2 q^{1 \over 8} \sin \pi u \cr
\Theta_2(u|\tau) \to 2 q^{1 \over 8} \cos \pi u \cr
\Theta_3(u|\tau) \to  \Theta_3(0|i \infty )=1 \cr
\Theta_4(u|\tau) \to  \Theta_4(0|i \infty )=1, \cr}$$
so that
$$\eqalign{
\lim_{q \to 0} Z_{2p+2}^{(D)}(u|\tau)&= {1 \over 2}
({\sin \pi (k+1)u \over \sin \pi u}
+{\cos \pi (k+1)u \over \cos \pi u}+2) \cr
&={\sin \pi ku \sin \pi ({k \over 2}+2)u \over
\sin \pi {k \over 2}u \sin 2 \pi u}. \cr}$$
Finally we use the lemma of appendix A of ref.\DY\ on
elliptic modular functions, to conclude that
the expression \dseries\ is identical to the D series elliptic genus
\eqn\ztwoii{Z_{2p+2}^{(D)}(u|\tau)=
{\Theta_1(ku|\tau) \Theta_1((k+4)u/2|\tau)
\over \Theta_1(ku/2|\tau) \Theta_1(2u|\tau)},}
as computed in \DY\ from the second potential \secondd, because the ratio
of both expressions is an elliptic modular form of weight zero, whose
$q \to 0$ limit equals one. Therefore, this ratio equals one identically.

\newsec{The $SU(3)$ case.}

\noindent{\bf Orbifolds and more for the $SU(2)$ factor.}

The $A$--type LG potentials for the $SU(3)$
KS models are generated by the function
\ref\GEPF{D. Gepner, Comm. Math. Phys. {\bf 141} (1991) 381.}
\eqn\potiii{\eqalign{
-\log(1-t \Phi_1 + t^2 \Phi_2)&=\sum_{m \geq 0}
t^m \ W_m^{(3)}(\Phi_1,\Phi_2) \cr
&=\sum_{m \geq 0} {t^m \over m}\
T_m(\Phi_1 / \Phi_2^{1/2}) \Phi_2^{m\over 2},\cr}}
where the index $m$ stands for $k+3$, $k$ the level of the $SU(3)$
factor of the KS model, and $T_m$ are the Chebyshev polynomials of
the first kind $T_m(2 \cos \theta)=2 \cos m \theta$.
These polynomials have interesting parity properties: the $T_{2m}$
are even, while the $T_{2m+1}$ are odd. Therefore, when $k+3$ is even,
the potential is a function of $\Phi_1^2$ and $\Phi_2$ only, say
$$W_{k+3}^{(3)}(\Phi_1,\Phi_2)=\Pi_{k+3}(\Phi_1^2,\Phi_2)$$
This fact allows moding out by the extra $\IZ_2$ symmetry already
encountered in the $SU(2)$ case of previous section. Setting
$\theta_1=\Phi_1^2$ and $\theta_2=\Phi_2$, and putting back the Jacobian
$\theta_1^{-1/2}$ into the action by introducing a third superfield
$\theta_3$, we get the $\IZ_2$ orbifold potential
\eqn\ztwoiii{ W_{k+3}^{(3,D)}(\theta_1,\theta_2,\theta_3)=
\Pi_{k+3}(\theta_1,\theta_2)+{1 \over 2}\theta_1 \theta_3^2.}
Following the lines of ref.\DY, we can perform the direct LG calculation
of the elliptic genus for this potential. Note that it
is quasi--homogeneous of degree $k+3$ for a grading of the fields
where $\theta_{1,2,3}$ have respective degrees $2,2,{k+1 \over 2}$.
The $U(1)$ transformation of the bosonic and fermionic components
of the fields
$\theta_i=\alpha_i+\theta_+ \beta_i^++\theta_- \beta_i^-$
which preserves the action reads
$$\eqalign{\alpha_1 &\to e^{2i\pi(2u)} \alpha_1 \cr
\alpha_2 &\to e^{2i\pi(2u)} \alpha_2 \cr
\alpha_3 &\to e^{2i\pi({k+1 \over 2}u)} \alpha_3 \cr
\beta_1^+ &\to e^{2i\pi(2u)} \beta_1^+ \cr
\beta_2^+ &\to e^{2i\pi(2u)} \beta_2^+ \cr
\beta_3^+ &\to e^{2i\pi({k+1 \over 2}u)} \beta_3^+ \cr
\beta_1^- &\to e^{-2i\pi(k+1)u} \beta_1^- \cr
\beta_2^- &\to e^{-2i\pi(k+1)u} \beta_2^- \cr
\beta_3^- &\to e^{-2i\pi({k+1 \over 2}+2)u} \beta_3^- \cr}$$
and the elliptic genus reads
\eqn\egd{ Z_3^{(D)}(u|\tau)= {\Theta_1(({k+1 \over 2}+2)u|\tau) \over
\Theta_1({k+1 \over 2}u|\tau)} \left({\Theta_1((k+1)u|\tau) \over
\Theta_1(2u|\tau)}\right)^2.}
Comparing this with the $A$--type elliptic genus \DY\
$$Z_3^{(A)}(u|\tau)={\Theta_1((k+1)u|\tau)\Theta_1((k+2)u|\tau) \over
\Theta_1(u|\tau)\Theta_1(2u|\tau)},$$
we see that the $\IZ_2$ orbifold has just replaced the
``$SU(2)_{k+1}$ factor" (with even level $k+1$)
of the $A$--type elliptic genus,
$\Theta_1((k+2)u|\tau)/\Theta_1(u|\tau)$, by the $\IZ_2$ orbifold
elliptic genus \ztwoii\ or equivalently \dseries, with $k \to k+1$.
This suggests that, although we do not know the corresponding LG potentials,
a similar mechanism should take place in the exceptional $E_{6,7,8}$
cases.
We conjecture that the corresponding elliptic genera are obtained
by replacing the ``$SU(2)_{k+1}$ factor" for $k+1=10,16,28$
by the corresponding $E_{6,7,8}$ elliptic genus \DY, namely
\eqn\conje{\eqalign{
Z_3^{(E_6)}(u|\tau)&=
{\Theta_1(10 u|\tau)\Theta_1(9u|\tau)\Theta_1(8u|\tau)
\over \Theta_1(2u|\tau)\Theta_1(3u|\tau)\Theta_1(4u|\tau)}\cr
Z_3^{(E_7)}(u|\tau)&=
{\Theta_1(16 u|\tau)\Theta_1(14u|\tau)\Theta_1(12u|\tau)
\over \Theta_1(2u|\tau)\Theta_1(4u|\tau)\Theta_1(6u|\tau)}\cr
Z_3^{(E_8)}(u|\tau)&=
{\Theta_1(28 u|\tau)\Theta_1(24u|\tau)\Theta_1(20u|\tau)
\over \Theta_1(2u|\tau)\Theta_1(6u|\tau)\Theta_1(10u|\tau)}.\cr}}
The corresponding potentials, if they exist, should be decorations
of the $E_{6,7,8}$ potentials involving an extra
field of dimension $2$. A possibility is given by the expressions
for the (flat, massive) perturbations of these theories
\ref\DVV{R. Dijkgraaf, E. Verlinde
and H. Verlinde, Nucl. Phys. {\bf B352} (1991) 59.}\
\ref\FLAT{W. Lerche, D. Smit and N. Warner, Nucl. Phys. {\bf B372}
(1992) 87;
S. Cecotti, Comm. Math. Phys. {\bf 131} (1990) 517 and
Comm. Math. Phys. {\bf 124} (1989) 23;S. Cecotti and C. Vafa,
Nucl. Phys. {\bf B367} (1991) 359; B. Blok and A. Varchenko,
Int. J. of Mod. Phys. {\bf A7} (1992) 1467;B. Dubrovin,
Nucl. Phys. {\bf B379} (1992) 627 and {\it Differential Geometry of the Space
of Orbits of a Coxeter Group}, Trieste preprint SISSA 29/93/FM (1993);
I. Krichever, Comm. Math. Phys. {\bf 143} (1992) 415;
S. Mahapatra, Phys. Lett. {\bf B264} (1991) 50;
A. Klemm, S. Theisen and M. Schmidt,
Int. J. of Mod. Phys. {\bf A7} (1992) 6215.},
which all involve a
``decoration" by a perturbation parameter of dimension two\foot{
see appendix B of \DLZ\ for
a recapitulation of these perturbed potentials. In this reference,
the dimension $2$ parameter is denoted by $t_{10,16,28}$
for $E_{6,7,8}$ respectively.}. The guess we make here is inspired by
the fact that the $SU(N)$ LG potentials for the KS models can be obtained
from the $SU(2)$ perturbed one, by a suitable substitution of the
dimensionful perturbation
parameters for LG by superfields \DLZ.
Moreover, the $D$ case treated above exhibits the same phenomenon:
the flat perturbations of the potential for the $D_{n+2}$ series
found in \DVV\
\DLZ\ are such that when retaining only
the dimension $2$ perturbation parameter (denoted by $t_{2n}$ in ref.\DLZ\ ),
it has exactly the form \ztwoiii\ with the substitution
$\theta_1 \to x$, $\theta_3 \to y$ and $\theta_2 \to t_{2n}$.
So we propose the following candidates, obtained by substituting
$t_{10,16,28} \to z$, a third superfield, in the expressions of
the corresponding E--type perturbed potentials. We get
\eqn\candid{\eqalign{
W_3^{(E_6)}(x,y,z)&={x^3 \over 3}+{y^4 \over 4}-x y^2 z
+{y^2 \over 2} z^3 +x{z^4 \over 12} \cr
W_3^{(E_7)}(x,y,z)&={x^3 \over 3}+xy^3
-3 x^2 y z +4 x^2 z^3 -3 xyz^4+x z^6
+{z^9 \over 6}\cr
W_3^{(E_8)}(x,y,z)&={x^3 \over 3}+{y^5 \over 5}-x y^3 z +xy^2z^4
-{6 \over 5}y^3 z^6 -{19 \over 15} xy z^7 +{28 \over 15}y^2 z^9\cr
&+{11 \over 45}x z^{10}-{82 \over 75} y z^{12}
+{103 \over 450} z^{15}.\cr}}
We checked that the LG elliptic genus for these potentials indeed
reproduces the above conjectures \conje.

Let us now compute the corresponding quantity in the framework of
the $N=2$ superconformal KS model.
The $A$ elliptic genus is by definition
a sum over Ramond characters
\eqn\racar{K_3(z|\tau)=\sum_{(\Gl_1,\Gl_2)\in P_k^{(3)}}
\chi_{\Gl_1,\Gl_1+2 \Gl_2}^{(\Gl_1,\Gl_2)}(z|\tau),}
where $P_k^{(3)}$ denotes the Weyl alc\^ove at level $k$ for
$SU(3)$ (``integrable weights", i.e. such that $\Gl_i\geq 0$ and
$\sum \Gl_i \leq k$), and the various indices on the Ramond character
of the KS coset denote from bottom to top respectively the $SU(2)$
weight,
$U(1)$ charge (chosen to be an integer) and $SU(3)$ weight.
The $\IZ_2$ orbifold is obtained by restricting the allowed $SU(2)$
weights $\Gl_1$ to the set
$Exp(D)=\{0,2,4,...,k+1\} \cup \{ {k+1 \over 2}\}$
corresponding to the allowed weights of the $SU(2)$ $D$ series.
Note that the weight $(\Gl_1=k+1, \Gl_2)$ is not in the
Weyl alc\^ove, but the Ramond character actually vanishes at this point,
so we can include it in the summation.
So we have the $\IZ_2$ orbifold elliptic genus
\eqn\ztoeg{K_3^{(D)}(z|\tau)=
\sum_{\Gl_1 \in Exp(D), \Gl_2\leq k-\Gl_1}
\chi_{\Gl_1,\Gl_1+2 \Gl_2}^{(\Gl_1,\Gl_2)}(z|\tau). }
We want to prove that this expression coincides with that of the
LG approach, eqn.\egd.
Using the explicit form of the characters and their modular transformations,
one can show that this function is a modular form of weight zero with
simple $z \to z+1$,$z\to z+\tau$ transformations
$$\eqalign{
K_3^{(D)}(z+1|\tau)&= K_3^{(D)}(z|\tau) \cr
K_3^{(D)}(z+\tau|\tau)&= e^{-18i\pi k(k+3)(2z+\tau)} K_3^{(D)}(z|\tau) \cr
K_3^{(D)}(z|\tau+1)&= K_3^{(D)}(z|\tau) \cr
K_3^{(D)}({z\over \tau}|-{1  \over \tau})&=
e^{18i\pi k(k+3){z^2 \over \tau}} K_3^{(D)}(z|\tau), \cr}$$
and with $q \to 0$ limit
$$\eqalign{\lim_{q \to 0}  K_3^{(D)}(z|\tau)&=
x^{-k}\sum_{\Gl_1 \in Exp(D), \Gl_2\leq k-\Gl_1} x^{\Gl_1+2 \Gl_2}\cr
&= x^{-k} {(1-x^{k+1})^2(1-x^{{k+1 \over 2}+2}) \over
(1-x^2)^2(1-x^{k+1 \over 2})},}$$
where we set $x=e^{6i\pi z}$. These transformations and limit coincide
with those of the LG result \egd\ provided we take $u=3z$. Hence, using the
lemma mentioned above on elliptic modular functions, we conclude that
$$ Z_3^{(D)}(u=3z|\tau)= K_3^{(D)}(z|\tau) .$$
The same calculation for the $E_{6,7,8}$ modular invariant theories
is tedious, but confirms the above
conjectures (where again we take $u=3z$). It involves in
particular a restriction of the sum over $\Gl_1$ to the sets of
shifted exponents of  $E_{6,7,8}$, namely
$$\eqalign{
Exp(E_6)&=\{ 0,3,4,6,7,10 \} \cr
Exp(E_7)&=\{ 0,4,6,8,10,12,16 \} \cr
Exp(E_8)&=\{ 0,6,10,12,16,18,22,28 \}, \cr }$$
which in particular yields the same $q \to 0$
limit as the above conjecture \conje, up to $u=3z$.
The last exponent $k+1=10,16,28$ is again outside of the Weyl alc\^ove,
but the characters vanish there, so we can safely
include them in the summation.
The only delicate point is the modular properties of the corresponding sums
$K_3^{(E)}$.
However, thanks to the general proof we gave in the introduction, the
modular covariance eqn. \transell\
is automatically ensured, with $\beta={18k(k+3)}$,
so the various transformations agree perfectly with those of \conje,
up to $u=3z$.
This is very strong evidence for the conjecture to actually be true.
It would be nice to derive our candidates for the LG potentials directly
from the $A$ potential $W_{k+3}^{(3)}$, although even in the $SU(2)$
case no direct link is known between the $E$--type potentials and the
$A$ ones.

\noindent{\bf $\IZ_3$ orbifold of the $SU(3)$ factor: the $\cal D$ series.}

The LG theory for the $A$--type $SU(3)$ KS model admits also a
$\IZ_3$ orbifold of the $SU(3)_{k}$ factor, whenever
$k$ is a multiple of $3$. The extra $\IZ_3$ symmetry of the potential
$W_{k+3}^{(3)}$ of eqn.\potiii\ for $3|k$ is readily seen to be the
{\it simultaneous} change
$$\Phi_1 \to e^{2i \pi l \over 3} \Phi_1 \qquad
\Phi_2 \to e^{4i \pi l \over 3} \Phi_2, $$
where $l \in \IZ_3$. So, we have to deal with an extra $\IZ_3$ twist
in our free field calculation for the LG elliptic genus, giving
rise to nine possible sectors according to the choice  of phases in
the $1$ and $\tau$ directions of the torus.
We choose to index these sectors by a fractional number
$a \in [-{1 \over 2},{1 \over 2})$, where in this case
$a \in \{ -{1 \over 3},0,
{1 \over 3} \}$, and to call $f_{a,b}(u|\tau)$ the elliptic genus
of the corresponding sector $(a,b) \in \{ -{1 \over 3},0,{1 \over 3} \}^2$.
The untwisted sector $a=b=0$ has the $A$--type elliptic genus
$$f_{0,0}(u|\tau)={\Theta_1((k+2)u|\tau) \over \Theta_1(u|\tau)}
\times {\Theta_1((k+1)u|\tau)\over \Theta_1(2u|\tau)},$$
where the two factors gather the respective contributions of the modes
of $\Phi_1$ and $\Phi_2$.
The elliptic genus $f_{a,b}(u|\tau)$ for the $(a,b)$ twisted sector is
easily obtained by a change $n \to n+a$ (resp. $n \to n+2a$) in the modes
of the components of $\Phi_1$ (resp. $\Phi_2$) for the $1$ direction,
and by the change
$u \to u+b$
for the $\tau$ direction.
For convenience, let us define
$$\Theta_{a,b}(u|\tau) = \sum_{n \in \IZ}
q^{\hlf (n+\hlf +a)^2}
e^{2i \pi (n+{1 \over 2}+a)(u+{1 \over 2}+b)} $$
which is proportional to $\Theta_1(u+a\tau +b|\tau)$,
satisfies $\Theta_{a+1,b}=\Theta_{a,b}$, and
which for $-1 < a \leq 0$ is equal to
$$  q^{{1 \over 2}(a+{1 \over 2})^2}
e^{2i \pi (a+{1 \over 2})(u+{1 \over 2}+b)}
(1-q^{-a}e^{-2i \pi (u+b)})
        \prod_{n \geq 1} (1-q^n)(1-q^{n+a}e^{2i \pi (u+b)})
(1-q^{n-a}e^{-2i\pi (u+b)}). $$
We recover the standard Jacobi theta functions for the choices
$$\eqalign{
\Theta_1\equiv \Theta_{0,0} &\qquad \Theta_2\equiv \Theta_{0,-{1 \over 2}}\cr
\Theta_3\equiv \Theta_{-{1 \over 2},-{1 \over 2}}
&\qquad \Theta_4\equiv \Theta_{-{1 \over 2},0}.\cr}$$
Let us also denote by $[x]$ the unique element of $x+\IZ$ in the interval
$[-{1 \over 2},{1 \over 2})$.
Then, the $(a,b)$ sector elliptic genus reads (for $k$ a multiple of $3$)
$$\eqalign{
f_{a,b}(u|\tau)&={\Theta_{[(k+2)a],[(k+2)b]}((k+2)u|\tau)
\Theta_{[(k+1)a],[(k+1)b]}((k+1)u|\tau) \over
\Theta_{a,b}(u|\tau)\Theta_{[2a],[2b]}(2u|\tau)}\cr
&={\Theta_{-a,-b}((k+2)u|\tau)
\Theta_{a,b}((k+1)u|\tau) \over
\Theta_{a,b}(u|\tau)\Theta_{-a,-b}(2u|\tau)},\cr}$$
where we used the fact that $[-x]=-[x]$ and $[3x]=0$
for $x \in \{0, \pm {1 \over 3} \}$.
    The total $\IZ_3$ orbifold elliptic genus is
\eqn\orbiii{ Z_3^{({\cal D})}(u|\tau)=
{1 \over 3} \sum_{a,b \in \{0, \pm {1 \over 3} \}} f_{a,b}(u|\tau).}
This result has again some particularly simple modular properties
thanks to the various transformations of $\Theta_{a,b}$
\eqn\tetprop{\eqalign{
\Theta_{a,b}(u+1|\tau)&= e^{2i \pi (a+{1 \over 2})} \Theta_{a,b}(u|\tau)\cr
\Theta_{a,b}(u+\tau|\tau)&=-e^{-i\pi(\tau+2u+2b)}\Theta_{a,b}(u|\tau)\cr
\Theta_{a,b}(u|\tau+1)&=e^{\ip ({1\over 4}-a^2)}\Theta_{a,a+b}(u|\tau)\cr
\Theta_{a,b}({u \over \tau}|-{1 \over \tau})&=(i \tau)^{1/2}
e^{2i\pi(a-{1 \over 2})(b+{1 \over 2})}
e^{i\pi{u^2 \over \tau}}\Theta_{b,-a}(u|\tau).\cr}}
To get the transformations of $f_{a,b}$, we still have to put the
resulting indices of $\Theta$ back into the interval
$[-{1 \over 2},{1 \over 2})$, at the possible cost of some phases.
In the $\IZ_3$ case here, we have $-[x]=[-x]$, therefore no phase appears
in the $(z,\tau)\to ({z \over \tau},-{1 \over \tau})$ transformation, but
in the $\tau \to \tau+1$ transformation,
we have to replace $a+b$ by $[a+b]=a+b-m$, introducing a phase
$e^{2i \pi (a+{1 \over 2})m}$. Remarkably, these phases cancel
each other in $f_{a,b}$, which transforms as
\eqn\ftran{\eqalign{
f_{a,b}(u+3|\tau)&=f_{a,b}(u|\tau)\cr
f_{a,b}(u+3\tau|\tau)&=e^{-6i\pi k(k+3) (3\tau+ 2u)} f_{a,b}(u|\tau)\cr
f_{a,b}(u|\tau+1)&=f_{a,a+b}(u|\tau)\cr
f_{a,b}({u \over \tau}|-{1 \over \tau})&=
e^{2i\pi k(k+3){u^2 \over \tau}} f_{b,-a}(u|\tau).\cr}}
Moreover, the $q \to 0$ limit of the LG orbifold elliptic genus is
easily derived, using the behaviors
\eqn\limits{\eqalign{
\Theta_{a,b}(u|\tau) &\to q^{{1 \over 2}(a+{1 \over 2})^2}
e^{2i \pi (a+{1 \over 2})(u+b+{1 \over 2})} \ \ \ {\rm for}\ \ a<0 \cr
&\to -2 q^{1 \over 8} \sin \pi (u+b) \ \ \ \ \ \ \
\ \ \ \ \ \ \ {\rm for} \ \ a=0 \cr
&\to q^{{1 \over 2}(a+{1 \over 2})^2-a}
e^{2i \pi (a-{1 \over 2})(u+b+{1 \over 2})} \ \ {\rm for}\ \ a>0,\cr}}
giving
$$\eqalign{
f_{a,b}(u|\tau) &\to  1 \ \ \ {\rm for}\ \ a \neq 0 \cr
&\to e^{-2i\pi k u} g(u+b) \ \
{\rm for} \ \ a=0,\cr}$$
with
$$g(u)={(1-e^{2i\pi (k+2)u})(1-e^{2i\pi (k+1)u})\over
(1-e^{2i\pi u})(1-e^{4i\pi u})},$$
so that the $q \to 0$ of the elliptic genus finally reads
$$\lim_{q \to 0}Z_3^{({\cal D})}(u|\tau)=
{1 \over 3}(e^{-2i\pi k u} [g(u)+g(u+{1 \over 3})
+g(u-{1 \over 3})]+6).$$

We now compare the LG orbifold elliptic genus to that of the
$\cal D$ series of the $SU(3)$ KS theories, corresponding to the
$\IZ_3$ orbifold of the $SU(3)$ factor.
The elliptic genus $K_3^{({\cal D})}(z|\tau)$
is obtained by restricting the sum of \racar\  to the $SU(3)$ weights
at level $k$ with triality zero ($\Gl_1 -\Gl_2 =0$
mod $3$), and with a triplication of the center of the Weyl alc\^ove
$({k \over 3},{k \over 3})$, the
fixed point of the $\IZ_3$ transformation.
It is again a straightforward exercise to derive the modular and
elliptic properties of the resulting sum, and we find that they match exactly
\ftran. Finally, the $q \to 0$ limit reads
$$x^{-k} \sum_{(\Gl_1,\Gl_2)\in P_k^{(3)} \atop
\Gl_1=\Gl_2 \ {\rm mod} \ 3}x^{\Gl_1+2\Gl_2} = {1 \over 3}
(x^{-k}[h(x)+h(\omega x)+h(\omega^2 x)] +6),$$
with $x=e^{6i\pi z}$, $\omega=e^{2i \pi \over 3}$, and
$$h(x)={(1-x^{k+1})(1-x^{k+2})\over (1-x)(1-x^2)}=g(u=3z).$$
Thanks to our lemma on elliptic modular functions, we conclude that
$$Z_3^{({\cal D})}(u=3z|\tau)=K_3^{({\cal D})}(z|\tau).$$

\newsec{The $\IZ_p$ orbifolds of the $SU(N)$ KS model.}

The LG potentials for the
$A$--type $SU(N)$ KS models are generated by \GEPF\
$$-\log(1-t\Phi_1+t^2 \Phi_2- \cdots +(-t)^{N-1} \Phi_{N-1})=
\sum_{m \geq 0} t^m W_m^{(N)}(\Phi_1,..,\Phi_{N-1}),$$
where $m$ stands for $k+N$, $k$ the level of the $SU(N)$ factor.
We can  form the $\IZ_p$ orbifold
of the LG theory at levels $k$ such that $p|k+N$.
In this case, the potential $W_{k+N}^{(N)}$ is invariant under the
simultaneous change
$$ \Phi_j \to e^{2i \pi l j \over p} \Phi_j \qquad j=1,..,N-1,$$
where $l \in \IZ_p$.
The corresponding twists introduce $p^2$ sectors $(a,b)$,
with $a,b \in \Sigma_p$, and
$$\eqalign{ \Sigma_p&= \{ 0,\pm {1 \over p}, \pm {2 \over p},\cdots,
\pm {p-1 \over 2p}\} \ \ \ {\rm for}\ \ p \ {\rm odd} \cr
&=\{-{1 \over 2},0,\pm {1 \over p}, \pm {2 \over p},\cdots,
\pm {p-2 \over 2p}\} \ \ {\rm for} \ \ p \ {\rm even}. \cr}$$
These correspond to the values of the twist of the LG free fields
in the $1$ and $\tau$ directions of the torus. Let $f_{a,b}(u|\tau)$
denote again the elliptic genus for the $(a,b)$ sector, $a,b \in \Sigma_p$.
The untwisted elliptic genus is that of the $A$ series computed in
\DY\ :
$$ f_{0,0}(u|\tau)=\prod_{j=1}^{N-1}
{\Theta_1((k+N-j)u|\tau) \over \Theta_1(ju|\tau)},$$
where the $j$-th factor in the product
gathers the contributions of the various modes of the superfield $\Phi_j$.
The effect of the twists $(a,b)$ is easily identified as
a shift of the modes $n \to n+a$ and of the $u$ variable $u \to u+b$,
up to a global phase, and we find
$$\eqalign{
f_{a,b}(u|\tau)&=\prod_{j=1}^{N-1}
{\Theta_{[(k+N-j)a],[(k+N-j)b]}((k+N-j)u|\tau) \over
\Theta_{[ja],[jb]}(ju|\tau) }\cr
&=\prod_{j=1}^{N-1}
{\Theta_{[-ja],[-jb]}((k+N-j)u|\tau) \over
\Theta_{[ja],[jb]}(ju|\tau) }.\cr}$$
    The total $\IZ_p$ orbifold elliptic genus reads
\eqn\orbpN{ Z_N^{(p)}(u|\tau) = {1 \over p} \sum_{a,b \in
\Sigma_p} f_{a,b}(u|\tau).}

To verify that this is correct, we need to check that we get
the same modular transformations and $q\to 0$ limit as can be derived
directly from the character expression for the elliptic genus.
This will give us conditions on what $p$'s are allowed (in addition to
the obvious condition $p|(k+N)$). Let us
start with the $q\to 0$ limit : we derive this for the general LG case
in the next chapter, and find an expression that differs from the
$q\to 0$ limit of the characters by a phase $(A_3)_{a,b}$, which must
therefore vanish (modulo 1)
for all $a,b \in \Sigma_p$. This phase is given (modulo 1) by
$$(A_3)_{a,b} = \sum_{[bj]\neq -\hlf} ( \sum_{[aj]\neq 0} (\hlf - aj) +
   \sum_{[aj]=0} bj ) $$
where the sums are over $j=1,...,N-1$. It is in fact simpler to work with
the difference $(A_3)_{a,b}-(A_3)_{a,0}$ which we will denote by
$B_{a,b}$ : it is given by
$$ B_{a,b} = \sum_{j=1}^{N-1} (\delta_{[aj]} - \delta_{[bj],-\hlf})
  (b-a)j. $$
We shall now show that this vanishes for all $a,b$ if and only if $p$
is a divisor of $N$ or of $N-1$ (meaning that $\IZ_p$ is a subgroup of
either the center of $G=SU(N)$ or of $H=SU(N-1)$). First, let us take
$a=0$ and $b={1\over p}$ : for this case we get
$$ B_{0,{1\over p}} = \sum_{j=1}^{N-1} (1-\delta_{[{j\over p}],-\hlf})
  ({1\over p})j = {1\over p}
\sum_{{j=1} \atop {[{j\over p}] \neq -\hlf}}^{N-1} j $$
and this must be an integer. Since the contribution of $j$'s for which
$[{j\over p}]=-\hlf$ to the sum is a multiple of $\hlf$ (a half integer),
we find that the
unrestricted sum ${1\over p}\sum_{j=1}^{N-1} j$ must be a half integer
as well, so that $p$ must be a divisor of $2\sum_{j=1}^{N-1}j=N(N-1)$.
Let us now take
$b={1\over p}$ with a general $a$ : we get
$$ B_{a,{1\over p}} = \sum_{j=1}^{N-1} (\delta_{[aj]}-
\delta_{[{j\over p}],-\hlf}) ({1\over p} - a) j $$
which must be an integer for all $a$.      We notice that once again
the contribution of terms with $[{j\over p}]=-\hlf$ to the sum is a
half integer, and so the expression
${1\over p}\sum_{j=1}^{N-1} j \delta_{[aj]}$ must also be a half integer,
meaning that $p$ must divide $2\sum_{j=1}^{N-1} j \delta_{[aj]}$ for all
$a$.      Since we found that $p|N(N-1)$, we can write
$p=qr$ where $q|N$ and $r|(N-1)$, and $q$ and $r$ have no common factors
since $N$ and $N-1$ are relatively prime.
Let us now take $a={1\over q}$ :
we get that $p$ must divide $2(q+2q+3q+...+(N-q))={{N(N-q)}\over q}$
but since $r$ and $N$ have no common factors this implies that $r$
divides $N-q$ and hence also that $r$ divides $q-1$ (since $r|(N-1)$).
Taking $a={1\over r}$ similarly gives that $p$ must divide
$2(r+2r+3r+...+(N-1))={{(N-1)(N-1+r)}\over r}$ so that $q$ must divide
$N-1+r$ and hence $q$ must divide $r-1$ (since $q|N$).
Since we found that $q|r-1$
and also $r|q-1$ either $r$ or $q$ must equal $1$, so that $p$ must
divide either $N$ or $N-1$ as claimed above. It is easy to check that
when this is satisfied, all the phases $(A_3)_{a,b}$ vanish, due to
cancellations between the contributions of the fields $\Phi_j$ and
$\Phi_{N-j}$.

The modular and elliptic transformations of the functions $f_{a,b}$
are derived from the properties of $\Theta_{a,b}$ \tetprop.
We find that
$$\eqalign{ f_{a,b}(u+N|\tau)&=e^{2\ip NA_N(a,b)}f_{a,b}(u|\tau)\cr
f_{a,b}(u+N\tau|\tau)&=e^{2\ip NB_N(a,b)}e^{-\ip kN(N-1)(k+N)(N\tau+2u)}
f_{a,b}(u|\tau)\cr
f_{a,b}(u|\tau+1)&=e^{2\ip C_N(a,b)} f_{a,a+b}(u|\tau)\cr
f_{a,b}({u \over \tau}|-{1 \over \tau})&=(i\tau)^{1/2}
e^{i \pi k(N-1)(k+N){u^2 \over \tau}}
e^{2\ip D_N(a,b)} f_{b,-a}(u|\tau),\cr}$$
where the phases $A_N,B_N,C_N,D_N$ combine the effect
of the phases in the transformations of the $\Theta$'s \tetprop\
and the necessity of putting back the indices of the transformed
$\Theta$'s into the interval $[-{1 \over 2},{1 \over 2})$.
The latter occurs only for the third and fourth transformations,
and gives contributions  due to
$$\eqalign{ \tau \to \tau+1
&: \Theta_{[a],[a]+[b]}(u|\tau)=
e^{2i \pi ([a]+{1 \over 2})([a]+[b]-[[a]+[b]])}
\Theta_{[a],[[a]+[b]]}(u|\tau) \cr
(u|\tau)\to ({u \over \tau}|-{1 \over \tau})
&:\Theta_{[b],-[a]}(u|\tau)= e^{2i\pi \delta_{[a],-{1 \over 2}}
([b]+{1 \over 2})} \Theta_{[b],[-a]}(u|\tau). \cr }$$
 We get by a simple calculation (with equalities satisfied modulo 1)
\eqn\trick{\eqalign{A_N(a,b)&= \sum_{j=1}^{N-1} [-ja]-[ja]
=\sum_{j=1}^{N-1} (-2ja) = -aN(N-1)   \cr}}
and
$$B_N(a,b)= \sum_{j=1}^{N-1} [jb]-[-jb] =-A_N(b,a),$$
so that both phases obviously vanish whenever $p|(N(N-1))$. The other
other phases come out to be
$$\eqalign{C_N(a,b)=&\sum_{j=1}^{N-1}
([-ja]+[-jb]-[-j(a+b)])([-ja]+{1 \over 2})-
([ja]+[jb]-[j(a+b)])([ja]+{1 \over 2}) \cr
=&\sum_{[ja]\neq -{1 \over 2}} [ja]([j(a+b)]+[-j(a+b)]-[jb]-[-jb])+\cr
&{1 \over 2}([j(a+b)]-[-j(a+b)]-[jb]+[-jb]-2[ja])\cr
=&\sum_{[bj]=-\hlf} (aj-\hlf) -
\sum_{[(a+b)j]=-\hlf} (aj-\hlf),\cr }$$
 and
$$\eqalign{D_N(a,b)&= \sum_{j=1}^{N-1}
([-ja]-{1 \over 2})([-jb]+{1 \over 2})
-([ja]-{1 \over 2})([jb]+{1  \over 2})+\delta_{[ja],-{1 \over 2}}
([-jb]-[jb])\cr
&=\sum_{[ja]\neq -{1 \over 2}} -[ja](1+[-jb]+[jb])+{1 \over 2}
([jb]-[-jb]) \cr
&=\sum_{[ja],[jb]\neq -\hlf} ([jb]-[ja])
=(b-a)\sum_{[ja],[jb]\neq -\hlf} j \cr}$$
and it is easy to check that all of these phases vanish whenever $p$ is
a divisor of $N$ or of $N-1$, again because of cancellations between the
contributions of $\Phi_j$ and $\Phi_{N-j}$.

The general $q\to 0$ limit is worked out in the next section.
As an example
 we can look at the case of $p=N$ with $N$ prime, for which we get
(using \limits)
$$\lim_{q \to 0} Z_N^{(p)}(u|\tau)={1 \over p} (x^{-k(N-1)/2}
[\sum_{j=0}^{p-1} h_N^{(k)}(\omega^j x) ]+p(p-1) ),$$
where $x=e^{2i \pi u}$, $\omega=e^{2i \pi \over p}$ (there are
only $p$ functions $f_{a,b}$ which have a limit different
from $1$, namely those with $a=0$), and
$$h_N^{(k)}(x)=\lim_{q \to 0}x^{k(N-1)/2} f_{0,0}(u|\tau)=
\prod_{j=1}^{N-1} {1-x^{k+N-j} \over 1-x^j}.$$

The Poincar\'e polynomial
of the corresponding ``chiral ring" $\IC[x_1,...,x_{N-1}]/
\nabla W_{k+N}^{(N)}$, namely
\eqn\recip{
P_N^{(p)}(x) = \sum_{{\rm ring}\ {\rm elements}} x^{\rm degree} , }
is equal to the $q\to 0$ limit of the elliptic genus (for $x$
defined as above), up to a constant power of $x$.
This polynomial must satisfy a duality symmetry
of the form
$$P(x)=x^m P({1 \over x}).$$
in any $N=2$ theory
\ref\CHF{ W. Lerche, C. Vafa and N. Warner, Nucl. Phys.
{\bf B324} (1989) 427.}\
\ref\GE{D. Gepner, Comm. Math. Phys. {\bf 142} (1991) 433.}\
with
$m=d\chat$ where $d$ is the degree of the Landau-Ginzburg
potential, and in our case $m=k(N-1)$.
This symmetry is assured
in our case whenever we have a modular invariant theory,
since the $x\to {1\over x}$ transformation is exactly the square
of the $S$ modular transformation.

\newsec{Orbifolds of general Landau-Ginzburg theories}

Let us consider now a general Landau-Ginzburg model, with a
quasi-homogenous potential including
fields of degrees (dimensions)
$d_i$ (for $i=1,...,n$) and a potential of degree $d$ (we
choose a normalization in which the degrees are all integers).
Classically, this theory has a $\IZ_d$ symmetry which takes the field
$\Phi_i$ to $e^{2\ip \omega {d_i \over d} } \Phi_i$ where $\omega$ is
a member of $\IZ_d$.
Obviously there is also a $\IZ_p$ symmetry for any $p$
which is a divisor of $d$. We shall now try
to orbifold the theory by such
a $\IZ_p$ factor - this means allowing the fields to be twisted by any
member of the symmetry group
when going around both non-trivial cycles of
the torus. The computation of the elliptic genus is reduced,
as in \WIT,
to a free field computation which is similar to the one performed in the
previous sections. The contribution to the elliptic genus
of the sector in which we have a
twist $a$ in one direction and a twist $b$ in the other
(meaning $\Phi_i$ transforms to $e^{2\ip ad_i}\Phi_i$ around one cycle
and to $e^{2\ip bd_i}\Phi_i$ around the other) comes out to be,
up to a constant phase related to the choice of the relative fermion
numbers of the vacua in the different sectors,
\eqn\eqfab{ \fab(\zt) =
\prod_j { {\Theta_{[-ad_j][-bd_j]}((d-d_j)z|\tau) } \over
  {\Theta_{[ad_j][bd_j]}(d_jz|\tau)} }. }
The total elliptic genus will then be
\eqn\kzt    { K(\zt) = {1\over p} \sum_{a,b=[0],[{1\over p}],...,
[{{p-1}\over p}]} e^{2\ip \pab} \fab(\zt).}
The phases $\pab$ will be determined later by the requirements of modular
invariance and a correct $q\to 0$ limit.
For now let us compute the modular transformations of $\fab$ : they come
out, using the transformations of the Theta functions \tetprop,
to be
$$ \fab(u+1|\tau) = (-1)^{nd}e^{-4\ip a\sum d_j} \fab(u|\tau) $$
$$ \fab(N(z+1)|\tau) = (-1)^{Nnd}e^{-4N\ip a\sum d_j} \fab(Nz|\tau) $$
$$ \fab(u+\tau|\tau)=e^{-\ip d^2 \chat(\tau+2u)}(-1)^{nd}\fab(u|\tau)$$
where as usual $\chat=\sum_j(1-{{2d_j}\over d})$, and
$$ \fab(N(z+\tau)|\tau)=e^{-\ip d^2N^2 \chat(\tau+2z)}
(-1)^{nNd}\fab(Nz|\tau).$$
We see that if we wish $u$ to be identified with $Nz$ (as we did for the
$SU(N)$ cases), modular invariance demands that $Nnd$ be even, and that
$p$ be a divisor of $2N\sum d_j$.

For the $\tau$ transformations we have :
$$ \fab(u|\tau +1)=e^{2\ip (A_1)_{a,b}}f_{a,a+b}(u|\tau) $$
where in the general orbifold
case there is a phase $(A_1)_{a,b}$ which arises from cases in
which $[ad_j]+[bd_j] \neq [(a+b)d_j]$ or $[-ad_j]+[-bd_j]\neq [-(a+b)d_j]
$, and which comes out to be (up to integer shifts)
$$ (A_1)_{a,b} = \sum_{[bd_j]=-\hlf} (ad_j - \hlf) -
         \sum_{[(a+b)d_j]=-\hlf} (ad_j - \hlf) $$
so that the condition we get for
modular invariance of $K(\zt)$ defined above
is that $(A_1)_{a,b}=\Phi_{a,a+b}-\pab$ (mod 1) for all $a,b$
(since $\fab$ satisfies $f_{a+1,b}=f_{a,b+1}=\fab$).
Note that for odd $r$ $(A_1)_{a,b}$ vanishes identically.

For the $S$ transformation we have :
$$\fab({u\over \tau}|-{1\over \tau})=e^{\ip d^2 \chat {u^2 \over \tau}}
    e^{2\ip (A_2)_{a,b}} f_{b,-a}(u|\tau) $$
where in this case the phase $(A_2)_{a,b}$
arises both from the $\tab$ transformation
and from corrections in case $-[a] \neq [-a]$. It comes out to be
$$ (A_2)_{a,b} =
\sum_j (([-ad_j]-\hlf)([-bd_j]+\hlf)-([ad_j]-\hlf)([bd_j]+\hlf)
      +\delta_{[ad_j],-\hlf} ([-bd_j]-[bd_j]))$$
which turns out to be equal (modulo 1) to
$$ (A_2)_{a,b} = \sum_{[ad_j]\neq -\hlf,[bd_j]\neq -\hlf} (b-a)d_j.$$
This phase should also satisfy $(A_2)_{a,b}=\Phi_{b,-a}-\pab$ (mod 1)
 for all $a,b$ for the elliptic genus to be modular invariant.
             Note, that if we act twice with this transformation we get
$$\fab(-u|\tau)=e^{2\ip ((A_2)_{a,b}+(A_2)_{b,-a})}
f_{-a,-b}(u|\tau)$$
and this transformation is exactly the duality transformation
 of the Poincar\'e polynomial ($x \to {1\over x}$)
related to the charge conjugation symmetry of $N=2$ theories.
We see that this polynomial is automatically
self-dual whenever our theory is modular invariant. The inverse of this
statement is not necessarily correct.

Let us now look at the $q\to 0$ limit of $\fab$.
 The contribution of each term of the form
$$ {{\Theta_{[-ad_j][-bd_j]}((d-d_j)z|\tau) } \over
  {\Theta_{[ad_j][bd_j]}(d_jz|\tau)} } $$
to this limit is one of the following :

(i) if $[ad_j]=-\hlf$ the contribution is 1.

(ii) if $-\hlf < [ad_j] < 0$ the contribution is
    $ e^{2\ip (-\hlf-[ad_j])(dz+1+[bd_j]+[-bd_j]) }.$

(iii) if $0 < [ad_j] < \hlf$ the contribution is
    $ e^{2\ip (\hlf-[ad_j])(dz+1+[bd_j]+[-bd_j]) }.$

(iv) if $[ad_j] = 0$ the contribution is
 $$ e^{-\ip ((d-2d_j)z+[-bd_j]-[bd_j]) }
    { {1 - e^{2\ip (z+b)(d-d_j)} } \over {1 - e^{2\ip (z+b)d_j} } }.$$

    The total contribution is therefore
$$P_{a,b}(x) = x^{d\sum_{[ad_j]\neq 0} (\hlf sign([ad_j]) - [ad_j])
               + \sum_{[ad_j]=0} (d_j - \hlf d) } e^{2\ip (A_3)_{a,b}}
        \prod_{[ad_j]=0} { {1 - (xe^{2\ip b})^{d-d_j} } \over
                           {1 - (xe^{2\ip b})^{d_j} } } $$
where $x=e^{2\ip z}$.
This expression was derived by a different method (and for $p=d$)
in \VAFORB.
Our expression differs from the result there and from the
direct free field computation (assuming zero fermion number for the
vacuum in all sectors) by a phase $(A_3)_{a,b}$,
which equals (modulo 1) to
$$ (A_3)_{a,b}=
\sum_{[ad_j]\neq 0}(\hlf sign([ad_j])-[ad_j])(1+[bd_j]+[-bd_j]) +
        \sum_{[ad_j]=0,[bd_j]\neq -\hlf} bd_j $$
or
$$(A_3)_{a,b}
= \sum_{[bd_j]\neq -\hlf} ( \sum_{[ad_j]\neq 0} (\hlf - ad_j) +
   \sum_{[ad_j]=0} bd_j ).$$

The only thing left in the computation is the determination of the phases
$\pab$. For the case of ${SU(N) \over \IZ_p}$ where $p$ is a divisor of
$N$ or of $N-1$ we checked
in previous chapters that all phase-corrections vanish and therefore we
can choose $\pab=0$ identically and get the correct elliptic genus.

In the general case, it is interesting to note that if we choose
\eqn\kztc{ K(\zt) = {1\over p} \sum_{a,b=[0],[{1\over p}],...,
[{{p-1}\over p}]} e^{-2\ip a\sum_j d_j}
              \prod_j{ {\Theta_{-[ad_j],-[bd_j]}((d-d_j)z|\tau) } \over
  {\Theta_{[ad_j][bd_j]}(d_jz|\tau)} }  }
we get no phases in the modular transformations (meaning the expression
is modular invariant), but we do get a phase of $\sum_{[ad_j]\neq 0}
(\hlf -2ad_j) + \sum_{[ad_j]=0} bd_j$ in the $q\to 0$ limit of the
$(a,b)$ sector.
If we want no phases in the $q\to 0$ limit, the phases $\pab$ are
uniquely determined to cancel the phases $(A_3)_{a,b}$, and we then
get conditions on the dimensions of the fields and on $p$ that our
theory must satisfy for the elliptic genus to be modular invariant.
Otherwise, there are many possible solutions $\pab$ to the modular
invariance conditions, giving different $q\to 0$ limits, so that some
condition must be externally imposed on the phases in this limit,
such as the one we got from the characters in the $SU(N)$ case.

\newsec{ Discussion and remarks}

In the present paper we discussed orbifolds associated with $N=2$
$G/H$ models which admit a LG description. Most of of our discussion is
related to the $SU(N)$ KS models. We proved that a consistent orbifold
theory is obtained
whenever we mode out by a $\IZ_p$ symmetry, with $p$ a divisor of $k+N$
and also of either $N$ or $N-1$.
This means that $\IZ_p$ is associated with a modular invariant of
either the $SU(N)$ or the $SU(N-1)$ factors of the
$SU(N)$ KS coset.
We calculated the elliptic genera associated with
these orbifolds using the LG potential of the $N=2$ superconformal
theory and
proved them to be equal to the elliptic genera of orbifolds of the $N=2$
SCFT.
In our approach we use the fact that the elliptic genus is modular
covariant.
Thus, whenever we can compute the elliptic genus starting from the LG
theory,
it is enough to check its modular properties and compare the $q\to 0$
limit
to that of the corresponding orbifold theory. If they match, the elliptic
genera match as well. This gives further support to the identification
of the macroscopic LG theory with the microscopic $N=2$ theory. The
orbifold theory, in
general, does not admit a LG description. Nevertheless, the LG potential of
the underlying $N=2$ SCFT encodes the information about the various orbifold
theories obtained by moding with respect to the appropriate $\IZ_p$
symmetry groups.

For example, in the $SU(3)$ KS case
we can construct not only the $\IZ_3$
orbifold associated with the $SU(3)$ factor in the coset, but also a
$\IZ_2$
orbifold associated with the $SU(2)$ factor of this coset. Our approach leads
us to conjecture that also some $E_{6,7,8}$ theories associated with the
corresponding $E$ modular invariants of the $SU(2)$ factor should exist.
We gave
a "natural" conjecture for their elliptic genera
and LG potentials based
on the
similarity between the perturbed $SU(2)$ potential and the unperturbed $SU(3)$
potential. It would be extremely interesting to establish whether one can
understand these theories also in terms of an orbifold construction. The
obvious suggestion that comes to mind is that (if at all) they can be
associated with the finite non-abelian (tetra, octa and icosa--edral)
subgroups of $SU(2)$.
Similar constructions can be employed for higher $SU(N)$ KS theories
taking advantage of the exceptional modular invariants. Again, it would be
tantalizing to conjecture that those may be associated with some non-abelian
orbifolds.

All these questions are intimately tied to the general question of the
relationship
between LG theories and CFT's. It is a challenging question to
investigate whether the LG theories can account for all CFT's via some definite
procedure. The orbifold approach, as demonstrated in this paper, allows for
the construction of new CFT's. It is an open problem to investigate
whether by allowing also for non--abelian
orbifolds we exhaust all modular
invariants associated with CFT's. Another, perhaps related, question is to
try to understand embeddings directly via the LG approach.

Although we were mainly interested in the $SU(N)$ KS theories, we did
investigate the elliptic genus associated with an orbifold of a general LG
theory. In the general case it is not clear exactly what the phase of the
$q\to 0$ limit of the elliptic genus should be, but for any definite
prescription for this phase we get conditions that must be satisfied
for a $\IZ_p$ orbifold to exist on the quantum level.

It would be extremely interesting to generalize our approach for the
construction of the elliptic genus also to other KS theories for which
the LG description is unknown (and may very well not exist). In many
cases the Poincar\'e polynomial, which is intimately
related to the elliptic genus, is known, and seems similar
(but not equal)
to a  polynomial connected to a LG theory of degree $k+h^v$
\GE\ \CHF\
(where $h^v$ is the second Casimir invariant of the group $G$). Thus we
may conjecture that these theories could be given by some orbifold of a
LG theory of this degree. Obviously for a $\IZ_p$ orbifold to exist in
this case $p$ must be a divisor of $k+h^v$, and based on our results here
it seems that $\IZ_p$ must also be a subgroup of either the center of $G$
or of the center of $H$ for modular invariance to be satisfied.
In particular it would be
important to clarify the relationship between the elliptic genera of
$N=2$ $G/H$
theories associated with a given $G$
and different choices for $H$.

Finally,
we would like to raise yet another open problem concerning the identification
of the macroscopic $N=2$ theory and the microscopic $N=2$ theories.
We have proven
that the elliptic genus and therefore the Ramond characters are encoded
within the LG potential. For a complete identification we would like also
to be able to get information pertaining to the characters of the NS sector
directly from the LG potential. The ultimate identification would be to
get all characters of the microscopic $N=2$ theory directly from the
corresponding LG potential.

\noindent{\bf Acknowledgements.}

We thank J.-B. Zuber for stimulating discussions
and a careful reading of the manuscript.

\noindent{\bf Note added.}

While this work was completed, there appeared on the hep-th net
a paper \ref\CONC{T. Kawai, Y. Yamada and S.-K. Yang,
{\it Elliptic Genera and N=2 Superconformal Field Theory},
KEK preprint 93/51 (1993).},
which slightly overlaps with some of our superconformal elliptic
genus results (SU(2) D series).

\listrefs
\end